\documentclass[twocolumn,showpacs,preprintnumbers,amsmath,amssymb]{revtex4}
\usepackage{graphicx}
\usepackage[english]{babel}
\usepackage{microtype}		
\usepackage{xcolor}	
\usepackage{mathtools}
\usepackage{braket}
\usepackage{tikz}
 \usepackage{printlen}

 \usepackage{longtable}
  \usepackage{booktabs}
 \usepackage{siunitx}
 \usepackage{csquotes}

\newcommand{\op}[1]{\ensuremath{\hat{#1}}}
\newcommand{\ladderdown}{\ensuremath{\op{a}^{\vphantom{\dagger}}}}
\newcommand{\ladderup}{\ensuremath{\op{a}^\dagger}}
\newcommand{\annihilop}{\ladderdown}
\newcommand{\creationop}{\ladderup}

\begin{document}
\bibliographystyle{apsrev}


\title{{\em Ab Initio} Quantum Monte Carlo Simulations of the Uniform Electron Gas \\ without Fixed Nodes}

\author{S.~Groth}
\author{T.~Schoof}
\author{T.~Dornheim}
\author{M.~Bonitz}
\affiliation{Institut f\"ur Theoretische Physik und Astrophysik, Christian-Albrechts-Universit\"{a}t zu Kiel, D-24098 Kiel, Germany}


\date{\today}

\begin{abstract}
The uniform electron gas (UEG) at finite temperature is of key relevance for many applications in the warm dense matter regime, e.g.\ dense plasmas and laser excited solids. Also, the quality of density functional theory calculations crucially relies on the availability of accurate data for the exchange-correlation energy. Recently, new benchmark results for the $N=33$ spin-polarized electrons at high density, $r_s={\bar r}/a_B \lesssim 4$ and low temperature, have been obtained with the configuration path integral Monte Carlo (CPIMC) method [T. Schoof \textit{et al.}, Phys. Rev. Lett. \textbf{115}, 130402 (2015)].
To achieve these results, the original CPIMC algorithm [T. Schoof \textit{et al.}, Contrib. Plasma Phys. \textbf{51}, 687 (2011)] had to be further optimized to cope with the fermion sign problem (FSP). It is the purpose of this paper to give detailed information on the manifestation of the FSP in CPIMC simulations of the UEG and to demonstrate how it can be turned into a controllable convergence problem. 
In addition, we present new thermodynamic results for higher temperatures. Finally, to overcome the limitations of CPIMC towards strong coupling, we invoke an independent 
method---the recently developed permutation blocking path integral Monte Carlo approach [T. Dornheim \textit{et al.}, accepted for publication in J. Chem Phys., arXiv:1508.03221]. 
The combination of both approaches is able to yield {\em ab initio} data for the UEG over the entire density range, above a temperature of about one half of the Fermi temperature. Comparison with restricted path integral Monte Carlo data [E. W. Brown \textit{et al.}, Phys. Rev. Lett. \textbf{110}, 146405 (2013)] allows us to quantify the systematic error arising from the free particle nodes.  
\end{abstract}

\maketitle
\section{Introduction\label{intro}}
The uniform electron gas (UEG) constitutes a well-known simple model for metals \cite{mahan}. At finite temperature, the spin-polarized UEG is described by the density parameter $r_s={\bar r}/a_B$ [${\bar r}$ is the mean interparticle distance related to the density by, $n^{-1}= 4\pi {\bar r}^3/3$, and $a_B$ is the Bohr radius] and the dimensionless temperature (degeneracy parameter) $\Theta= k_BT/E_F$, with the Fermi energy $E_F$. Besides being an interesting theoretical model system for studying correlated fermionic many-body systems, exact data for the exchange-correlation energy of the UEG is essential for the construction of exchange correlation functionals \cite{karasiev_prl14, brown_prb13}, for density functional theory (DFT) calculations of more realistic systems, e.g.\ atoms, molecules and novel materials. For the ground state this data has been provided many years ago by Ceperley and Alder \cite{ceperley_alder} utilizing the fixed node diffusion Monte Carlo approach. Based on these calculations Perdew and Zunger computed the density functionals \cite{perdew_zunger}, which have been the basis for countless DFT applications. 

Often one is interested in properties of chemical systems or condensed matter at low temperature, not exceeding room-temperature, for which it is justified to use ground state results. 
However, in recent years more and more applications have emerged where the electrons are highly excited, e.g.\ by compression of the material or by electromagnetic radiation. Examples are dense plasmas in compact stars or planet cores, e.g.\ \cite{knudson_12, Militzer:2008, Nettelmann:2012} and laser fusion experiments at the National Ignition Facility, e.g.\ \cite{lindl_04, hu_11, hurricane_nif14}, at Rochester \cite{theobald_15} or Sandia~\cite{hanson_14,hanson_14_2}. It is now widely agreed upon that the theoretical description of these experiments requires to go beyond ground state DFT. This leads to a high demand for exact data for the UEG at finite temperature and high to moderate density where fermionic exchange and correlation effects play an important role simultaneously, namely the warm dense matter (WDM) regime, where both $r_s$ and $\Theta$ are of order one.  
\par

Quantum Monte Carlo (QMC) simulations are the method of choice for the computation of exact thermodynamic quantities at finite temperature. However, it is well-known that, when applied to fermions, path integral Monte Carlo (PIMC) methods suffer the fermion sign problem (FSP), which may render the simulation even of small fermionic systems impossible and was shown to be NP-hard~\cite{troyer}. In the standard PIMC formulation in coordinate space, e.g.~\cite{ceperley95rmp}, the FSP causes an exponential loss of accuracy with increasing degeneracy, i.e.~towards low temperature and high density of the system. For this reason, standard fermionic PIMC calculations of the $N=33$ spin-polarized UEG are not feasible in the warm dense matter regime~\cite{brown_prl13}. Presently, the search for accurate and efficient strategies to weaken the FSP is one of the most important questions in condensed matter and dense plasma theory. 

A popular approach to avoid the FSP is the restricted (fixed-node) PIMC (RPIMC) method~\cite{ceperley_91}, which is claimed to be exact if the true nodal surface of the density matrix would be known. Usually this is not the case, and one has to rely on approximations, thereby introducing an uncontrolled systematic error.
Brown \textit{et al.}~\cite{brown_prl13} performed RPIMC calculations with ideal nodes of the UEG in a broad density-temperature range down to $r_s=1$ and $\Theta=0.0625$. These results have been used by many groups, e.g. for the construction of analytical fits for the exchange-correlation free energy \cite{karasiev_prl14, brown_prb13} and as benchmarks for models and simulations \cite{filinov15, dubois14}.

In a recent Letter~\cite{tim_prl15}, we applied the configuration path integral Monte Carlo (CPIMC) approach to the uniform electron gas and were able to obtain the first {\em ab initio} simulation results for finite temperatures and high degeneracy. These results also showed that the RPIMC data of Ref.~\cite{brown_prl13} are inaccurate for high densities, $r_s \lesssim 4$. As any fermionic PIMC approach, CPIMC as well suffers from the FSP. But, 
being 
 formulated in Fock space of Slater determinants~\cite{schoof_cpp_11, schoof_cpp14}, CPIMC experiences an increasing FSP with decreasing quantum degeneracy, i.e. towards low density. In the case of the UEG with $N=33$ particles, direct CPIMC simulations were possible only for $r_s \lesssim 0.4$. Nevertheless, in Ref.~\cite{tim_prl15} an extension to substantially larger $r_s$ was achieved by introducing an auxiliary kink potential which leads to a complication of the original CPIMC algorithm.
 
\begin{figure}
 \includegraphics[width=85mm]{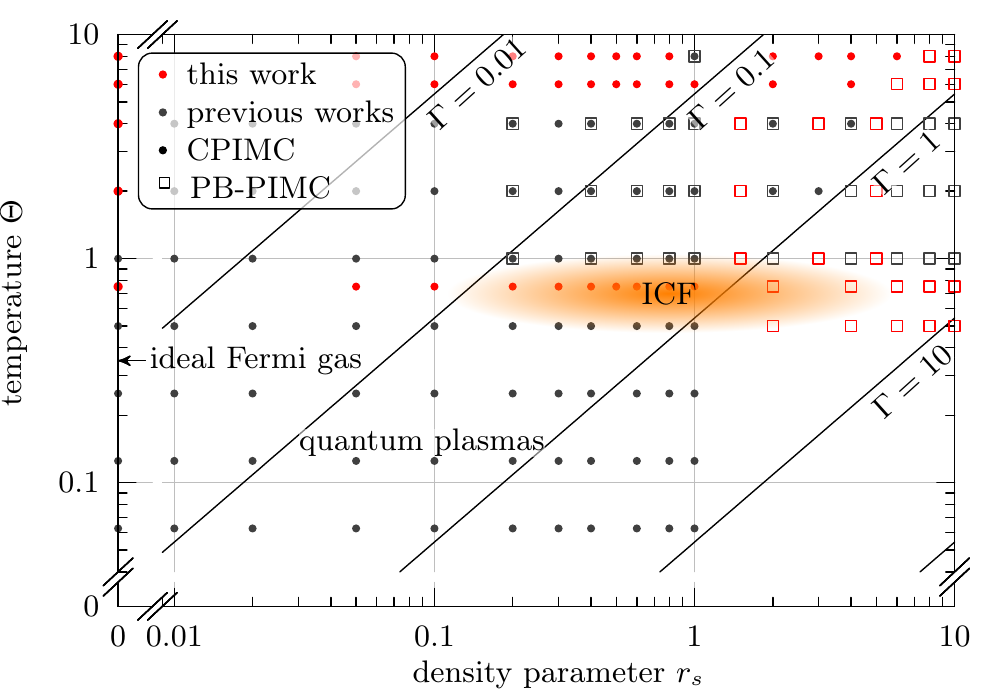}
 \caption{Available {\em ab initio}-Quantum Monte Carlo data in the  warm dense matter range for $N=33$ spin-polarized electrons. Dots: CPIMC. Squares: PB-PIMC.
 Red: Additional combined CPIMC and PB-PIMC results of this work. Gray: Previous results from CPIMC~\cite{tim_prl15} and PB-PIMC~\cite{pb_pimc_jcp15}, respectively. 
ICF: typical inertial confinement fusion parameters \cite{hu_11}. Quantum (classical) behavior dominates below (above) the line $\Theta=1$. $\Gamma=e^2/{\bar r}k_BT$  is the classical coupling parameter.}
 \label{fig:n-t-sketch}
\end{figure}

For this reason, the present article aims at giving a comprehensive explanation of the modified CPIMC approach, in particular, of the details of the kink potential and the issues of convergence and accuracy. In order to give a systematic analysis of these concepts and their capabilities, we concentrate on the simplest situation---the polarized UEG. Also, we restrict ourselves to finite particle numbers, deferring the issues of finite size effects and extrapolation to the thermodynamics limit to a future publication. Here, we explore in detail how the algorithm performs with varying particle number and what range of densities and temperatures is accessible. This allows us to extend the range of {\em ab initio} CPIMC data presented in Ref.~\cite{tim_prl15} to temperatures as high as $\Theta = 8$ and to larger $r_s$-values, where the maximum accessible value is found to be on the order of $r^{\rm max}_{s} \sim \Theta$. However, we demonstrate that it is possible to access the  entire $r_s$-range without fixed nodes. To this end, we invoke another {\em ab initio} approach---the recently developed permutation blocking PIMC method (PB-PIMC)~\cite{pb_pimc_njp, pb_pimc_jcp15} which has a complementary FSP, restricting the simulations from the side of low temperatures.
For $N=33$ spin-polarized particles, the combination of CPIMC and PB-PIMC allows us to present exact results for $\Theta \ge 0.5$, for all densities, without fixed nodes, see Fig.~\ref{fig:n-t-sketch}.



The paper is organized as follows.
After introducing the model Hamiltonian of the UEG in Sec.~\ref{sec:model}, we start with a brief but self-contained derivation of the CPIMC expansion of the partition function in Sec.~\ref{sec:expansion} and, in Sec.~\ref{sec:path}, explain the interpretation of the latter as being a sum over closed paths in Fock space, in imaginary time. In Sec.~\ref{sec:sign_direct}, we proceed with addressing the FSP in direct CPIMC simulations, where we find an abrupt drop of the average sign at a certain critical value of $r_s$ depending on particle number and temperature. Then, in Sec.~\ref{sec:kinkPot}, we demonstrate how the applicable region of the CPIMC method can be extended to significantly lower densities by the use of an auxiliary kink potential and an appropriate extrapolation scheme. In Sec.~\ref{sec:PB-PIMC}, the main ideas of PB-PIMC and its differences compared to standard PIMC are explained. Finally, in Sec.~\ref{sec:results}, we combine the two complementary methods, CPIMC and PB-PIMC, for the first time to obtain benchmark results for $N=33$ spin-polarized particles over the whole density range for several degeneracy parameters reaching from $\theta =0.5$ to $\theta=8$.

\section{Theory}
\subsection{The Jellium Hamiltonian \label{sec:model}} 
In second quantization with respect to plane waves,
$\langle \mathbf{r}\;|\mathbf{k}\rangle = \frac{1}{L^{3/2}} e^{i\mathbf{k} \cdot \mathbf{r}}$ with $\mathbf{k}=\frac{2\pi}{L}\mathbf{m}$, $\mathbf{m}\in \mathbb{Z}^3$, 
the Hamiltonian of the finite simulation-cell 3D uniform electron gas consisting of $N$ electrons on a uniform neutralizing background in a periodic box of length $L$ takes the familiar form (Rydberg units)
%
\begin{align}\label{eq:h} 
& \op{H} =
\sum_{i}\mathbf{k}_i^2 \creationop_{i}\annihilop_{i} + 2\smashoperator{\sum_{\substack{i<j,k<l \\ i\neq k,j\neq l}}} 
w^-_{ijkl}\creationop_{i}\creationop_{j} \annihilop_{l} \annihilop_{k} + E_M,
\end{align}
with the antisymmetrized two-electron integrals, $w^-_{ijkl} =w_{ijkl}-w_{ijlk}$, where
\begin{align} 
\quad w_{ijkl}=\frac{4\pi e^2}{L^3 (\mathbf{k}_{i} - \mathbf{k}_{k})^2}\delta_{\mathbf{k}_i+\mathbf{k}_j, \mathbf{k}_k + \mathbf{k}_l}\;,
\label{eq:two_ints}
\end{align}
and the delta-function ensuring momentum conservation. The first (second) term in the Hamiltonian Eq.~(\ref{eq:h}) describes the kinetic (interaction) energy. The Madelung energy $E_M$ accounts for the self-interaction of the Ewald summation in periodic boundary conditions~\cite{fraser_madelung}, for which we found $E_M\approx-2.837297\cdot(3/4\pi)^{\frac{1}{3}}N^{\frac{2}{3}}r_s^{-1}$. 
The operator 
$\creationop_{i}$  ($ \annihilop_{i}$) 
creates (annihilates) a particle in the orbital $|\mathbf{k}_i\rangle$.
The diverging contributions in the interaction term, i.e.\ for $\mathbf{k}_i=\mathbf{k}_k$ and $\mathbf{k}_j=\mathbf{k}_l$, cancel with the contributions due to the positive background. 
Note that choosing the plane wave basis, which is the ideal, natural and Hartree-Fock basis at the same time, has the major advantage of having two-electron integrals that can be computed analytically according to Eq.~(\ref{eq:two_ints}). In an arbitrary basis one generally has to compute the two-electron integrals prior to the simulation and store them in computer memory, limiting the number of basis functions that can be taken into account. Yet, it is well-known that plane waves badly describe the Coulomb interaction, making a large number of basis functions necessary to obtain converged results.  

\subsection{CPIMC Expansion of the partition function \label{sec:expansion}}

In equilibrium many-body quantum statistics the central quantity is the partition function, which is given by the trace over the density operator
\begin{align}
Z=\text{Tr} \;{\op{\rho}}\;,
\label{eq:Z_trace}
\end{align}
where, in the canonical ensemble,
\begin{align}
\op{\rho} = e^{-\beta\op{H}},
\label{eq:rho}
\end{align}  
with the inverse temperature $\beta=[k_BT]^{-1}$. In standard PIMC, the trace in Eq.~(\ref{eq:Z_trace}) is evaluated in coordinate space expressing the density operator in terms of a product of $M$ density operators at $M$-times higher temperature, which is justified by the Trotter formula. To correctly take into account Fermi statistics, one then has to antisymmetrize the density operator thereby introducing a sign change in the weight function for odd particle permutations. This is the source of the FSP in standard PIMC. In CPIMC instead we perform the trace in Eq.~(\ref{eq:Z_trace}) directly with antisymmetrized $N$-particle states (Slater determinants)
\begin{align}
|\{n\}\rangle=|n_1, n_2, \dots\rangle\;,
\end{align}  
which form a complete basis of the Fock space. Here, the $n_i$ denote the fermionic occupation numbers ($n_i=0, 1$) of the orbitals $|\mathbf{k}_i\rangle$. 

To bring the partition function into a form suitable for a Monte Carlo algorithm, one can split the Hamiltonian into a diagonal and off-diagonal part, i.e.\ $\op{H}=\op{D}+\op{Y}$, which is always possible for any arbitrary basis. In the interaction picture in imaginary time with respect to the diagonal operator $\op{D}$, i.e.
\begin{align}
\op{Y}(\tau)&=e^{\tau\op{D}}\op{Y}e^{-\tau\op{D}},\;\; \tau\in(0,\beta)\;,
\end{align}
the density operator can be written in terms of a perturbation expansion in orders of $\op{Y}$
\begin{align}
  e^{-\beta\op{H}}&=e^{-\beta\op{D}}\op{T}_\tau e^{-\int_0^\beta\op{Y}(\tau)\mathrm{d}\tau}\nonumber\\ 
&=e^{-\beta\op{D}}\sum_{K=0}^\infty \int\limits_{0}^{\beta} d\tau_1 \int\limits_{\tau_1}^{\beta} d\tau_2 \ldots \int\limits_{\tau_{K-1}}^\beta d\tau_K\nonumber\\  
&\qquad\qquad(-1)^K\op{Y}(\tau_K)\op{Y}(\tau_{K-1})\cdot\ldots\cdot\op{Y}(\tau_1)\;,
\label{eq:identity}
\end{align}
where $\op{T}_\tau$ denotes the time-ordering operator. Inserting Eq.~(\ref{eq:identity}) into Eq.~(\ref{eq:Z_trace}), evaluating the trace and rearranging terms yields the following expansion of the partition function
\begin{align}
Z = &
\sum_{K=0,\atop K \neq 1}^{\infty} \sum_{\{n\}}
\sum_{s_1\ldots s_{K-1}}\,
\int\limits_{0}^{\beta} d\tau_1 \int\limits_{\tau_1}^{\beta} d\tau_2 \ldots \int\limits_{\tau_{K-1}}^\beta d\tau_K 
\label{eq:Z_expansion} \\\nonumber
& (-1)^K  
e^{-\sum\limits_{i=0}^{K} D_{\{n^{(i)}\}} \left(\tau_{i+1}-\tau_i\right) } 
\prod_{i=1}^{K} Y_{\{n^{(i)}\},\{n^{(i-1)}\} }(s_i)\;,
\end{align}
where $s_i$ denotes a multi-index defining the orbitals in which the two sets of occupation numbers $\{n^{(i)}\}$ and  $\{n^{(i-1)}\}$ differ. Due to the trace in Eq.~(\ref{eq:Z_trace}) it has to be  $\{n\}=\{n^{(0)}\}=\{n^{(K)}\}$. According to the Slater-Condon rules the Fock space matrix elements of the UEG Hamiltonian do not vanish only if the states differ in no (diagonal part) or exactly four occupation numbers (off-diagonal part) so that  
\begin{align}
 &D_{\{n^{(i)}\}} = \sum_l \mathbf{k}_l^2 n^{(i)}_{l} + \sum_{l<k}w^-_{lklk}n^{(i)}_{l}n^{(i)}_{k} \;,\label{eq:diagonal}\\
&Y_{ \{n^{(i)}\},\{n^{(i-1)}\} }(s_i) =w^-_{s_i}(-1)^{\alpha^{\phantom{-}}_{s_i}}\; 
\label{eq:off_diagonal}
\end{align}
with $s_i=(pqrs)$ defining the four occupation numbers in which  $\{n^{(i)}\}$ and $\{n^{(i-1)}\}$ differ, where it is $p<q$ and $r<s$. In this notation, the exponent of the fermionic phase factor is given by  
\begin{align}
\alpha^{\phantom{-}}_{s_i} &=\alpha^{(i)}_{pqrs}=\sum_{l=p}^{q-1}n^{(i-1)}_{l}+\sum_{l=r}^{s-1}n^{(i)}_{l}\;.
\nonumber
\end{align}
Monte Carlo estimators of observables are readily computed as derivatives of the partition function Eq.~(\ref{eq:Z_expansion}),
e.g. for the internal energy one obtains
\begin{align}
\langle\op{H}\rangle=& - \frac{\partial}{\partial \beta} \ln Z \\
=&
\sum_{K=0,\atop K \neq 1}^{\infty} \sum_{\{n\}}
\sum_{s_1\ldots s_{K-1}}\,
\int\limits_{0}^{\beta} d\tau_1 \int\limits_{\tau_1}^{\beta} d\tau_2 \ldots \int\limits_{\tau_{K-1}}^\beta d\tau_K \nonumber\\ 
&\biggl(\frac{1}{\beta} \sum_{i=0}^K D_{\{n^{(i)}\}}(\tau_{i+1}-\tau_i) -\frac{K}{\beta}\biggr) W\;. \label{eq:energy_estimator}
\end{align}
We point out that the expansion~(\ref{eq:Z_expansion}) is exact and system-independent. Monte Carlo methods using this expansion belong to the so-called continuous time QMC methods (in the interaction picture) since there is no imaginary time discretization left. This concept has been developed by Prokof’ev \textit{et al.}~\cite{prokofiev96,prokofiev98} and extensively applied to lattice models, e.g.~\cite{prokofiev96,prokofiev98,houcke06,rombouts06}. We have presented an alternative derivation of Eq.~(\ref{eq:Z_expansion}) by starting from the Trotter formula and developed an algorithm for continuous systems~\cite{schoof_cpp_11} requiring more involved Monte Carlo steps compared to lattice models.

\subsection{Closed path in Fock space\label{sec:path}}
\begin{figure}
 \begin{tikzpicture}[xscale=0.74, yscale=0.4]
\newcommand{\xrange}{10}
\newcommand{\yrange}{1}
\newcommand{\taueins}{\xrange*0.3}
\newcommand{\tauzwei}{\xrange*0.45}
\newcommand{\taudrei}{\xrange*0.8}
\newcommand{\zero}{\yrange*1}
\newcommand{\eins}{\yrange*2}
\newcommand{\zwei}{\yrange*3}
\newcommand{\drei}{\yrange*4}
\newcommand{\vier}{\yrange*5}
\newcommand{\fuenf}{\yrange*6}

\draw[->] (0,0) -- +(\xrange+0.5*\xrange/20,0) coordinate (xlabel);
\draw[->] (0,0) -- +(0,\yrange*7) coordinate (ylabel);
\foreach \i in {0,...,5} {
	\draw (-0.1,\i*\yrange+\yrange) node[left] {$\i$} -- (0.1,\i*\yrange+\yrange);
}
\foreach \i/\l in {\taueins/$\tau_1$,\tauzwei/$\tau_2$,\taudrei/$\tau_3$} {
\draw (\i,0.1) -- (\i,-0.1) node[below] {\l};
}
\draw (0,0.1) -- (0,-0.1) node[below] {$0$};
\draw (\xrange,0.1) -- (\xrange,-0.1) node[below] {$\beta$};
\node at (0.5*\xrange,-1.5) {imaginary time $\tau$};
\node[rotate=90] at (-0.8,3.5*\yrange) {orbital $i$};

\foreach \i in {1,...,6} {
\draw[semithick,dotted] (0,\i*\yrange) -- (\xrange,\i*\yrange);
}

\draw[dashed] (\tauzwei+2.0,0.5) rectangle (\tauzwei+2.75,\yrange*6.5);
\node at (\tauzwei + 3.0,\fuenf+\yrange+0.2) {$|\{n^{(2)}\}\rangle=|001110\ldots\rangle$};

\begin{scope}[thick]
\draw (0,\zero) -| (\taueins,\zwei) -| (\taudrei,\zero) -- (\xrange,\zero);
\draw (0,\eins) -| (\tauzwei,\vier) -| (\taudrei,\eins) -- (\xrange,\eins);
\draw (0,\drei) -| (\taueins,\fuenf) -| (\tauzwei,\drei) -- (\xrange,\drei);
\draw (\taueins, \drei) -- (\taueins, \zwei);
 \end{scope}
 
 \node at (\taueins, \fuenf+\yrange+0.2) { $s_1 =(2,5,0,3)$};

\end{tikzpicture}
 \caption{Typical closed path in Slater determinant (Fock) space. The state with three occupied orbitals $|{\vec k}_0\rangle, |{\vec k}_1\rangle, |{\vec k}_3\rangle$ undergoes a two-particle excitation $s_1$ at time $\tau_1$ replacing the occupied orbitals $|{\vec k}_0\rangle, |{\vec k}_3\rangle$ by 
$|{\vec k}_2\rangle, |{\vec k}_5\rangle$. Two further excitations occur at $\tau_2$ and $\tau_3$.
The states at the ``imaginary times'' $\tau = 0$ and $\tau = \beta$ coincide. All possible paths contribute to the partition function $Z$, Eq.~(\ref{eq:Z_expansion}). (Fig.~from \cite{tim_prl15})}
 \label{fig:sketch}
\end{figure}
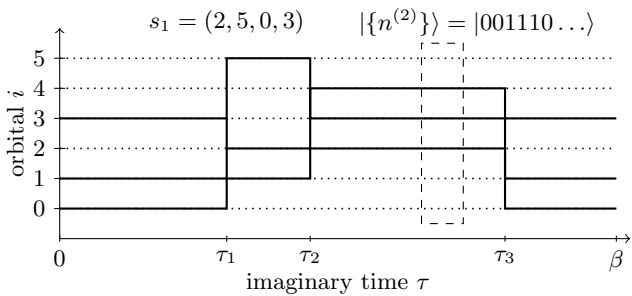
A contribution to the partition function Eq.~(\ref{eq:Z_expansion}) can be interpreted as a $\beta-$periodic path in Fock space, in imaginary time, that is uniquely defined by the initial determinant $\{n\}=\{n^{(0)}\}$ at $\beta=0$ and the $K$ two-particle excitations of type $s_i=(pqrs)$ at times $\tau_i$, where two particles are excited from the orbitals $r$ and $s$ to $p$ and $q$. An example of such a path is illustrated in Fig.~\ref{fig:sketch}. Due to their visual appearance, the excitations are called ``kinks''. The weight of each path is determined by the weight function which, according to Eq.~(\ref{eq:Z_expansion}) and~(\ref{eq:off_diagonal}), reads
\begin{align}
& W(K,\{n\}, s_1,\ldots,s_{K-1},\tau_1,\ldots,\tau_K)=\label{eq:weight}\;\\ \nonumber
&\qquad\qquad (-1)^K  
e^{-\sum\limits_{i=0}^{K} D_{\{n^{(i)}\}} \left(\tau_{i+1}-\tau_i\right) } 
\prod_{i=1}^{K} w^-_{s_i}(-1)^{\alpha^{\phantom{-}}_{s_i}}\;.
\end{align} 
The set of occupation numbers of a determinant between kinks contributes to the exponential function with its corresponding diagonal matrix element, cf.\ Eq.~(\ref{eq:diagonal}), weighted with the length of the time interval on which the determinant is realized in the path. On the other hand, each kink enters the product over all kinks in the path with its corresponding antisymmetrized two-electron integral and phase factor of the involved orbitals. Since the two-electron integrals can be both positive and negative, there are altogether three sources of sign changes in the weight function.
 
\section{Sign problem of CPIMC \label{sec:sign_in_CPIMC}}
\subsection{Sign problem of the direct CPIMC method \label{sec:sign_direct}}
Since the weight function $W$ takes both positive and negative values, it is not a probability density. Therefore, the Metropolis algorithm can only be used to generate a Markov chain of paths distributed according to the modulus of the weight. This is achieved with an ergodic set of six Monte Carlo steps in which single or paired kinks are added or changed. A detailed description of these steps can be found in~\cite{schoof_cpp14}. By generating a Markov chain of paths according to the modulus of the weight, we actually simulate a system described by
\begin{align}
Z'=& \sum_{K=0,\atop K \neq 1}^{\infty} \sum_{\{n\}} \sum_{s_1\ldots s_{K-1}}\int\limits_{0}^{\beta} d\tau_1 \int\limits_{\tau_1}^{\beta} d\tau_2 \ldots \int\limits_{\tau_{K-1}}^\beta d\tau_K\,\label{eq:primed_system} \\ \nonumber 
&\qquad\qquad |W(K,\{n\}, s_1,\ldots,s_{K-1},\tau_1,\ldots,\tau_K)| \;,
\end{align} 
rather than the true physical system described by the partition function Eq.~(\ref{eq:Z_expansion}). Physical expectation values of observables are then obtained via
\begin{align}
\langle O \rangle = \frac{\langle Os\rangle^\prime}{\langle s \rangle^\prime}\;, 
\label{eq:average}
\end{align}
where $O$ is the Monte Carlo estimator, e.g.\ for the internal energy the term in brackets in Eq.~(\ref{eq:energy_estimator}), $\langle\cdot\rangle^\prime$ denotes the expectation value with respect to the modified partition function, Eq.~(\ref{eq:primed_system}), and $s=\text{sign} (W)$ measures the sign of each path. For the expectation value of $s$, which is called the \emph{average sign}, it holds
\begin{align}
\braket{s}'= \frac{Z}{Z'}=e^{-\beta N(f-f')}
\end{align}
with f being the free energy per particle. It is straight forward to show that the relative statistical error of quantities computed with Monte Carlo methods via Eq.~(\ref{eq:average}) is inversely proportional to the average sign. Therefore, it grows exponentially with particle number and inverse temperature while it can only be reduced by the square root of the number of Monte Carlo samples. Depending on the available computational resources acceptable statistical errors can be obtained for average signs larger than about $10^{-4}$. This is the FSP.

\begin{figure}[t]
\includegraphics{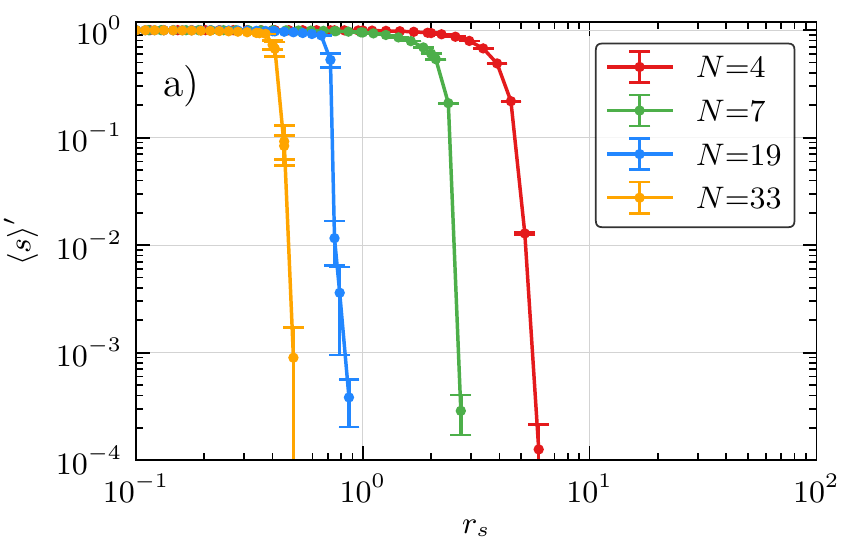}
\includegraphics{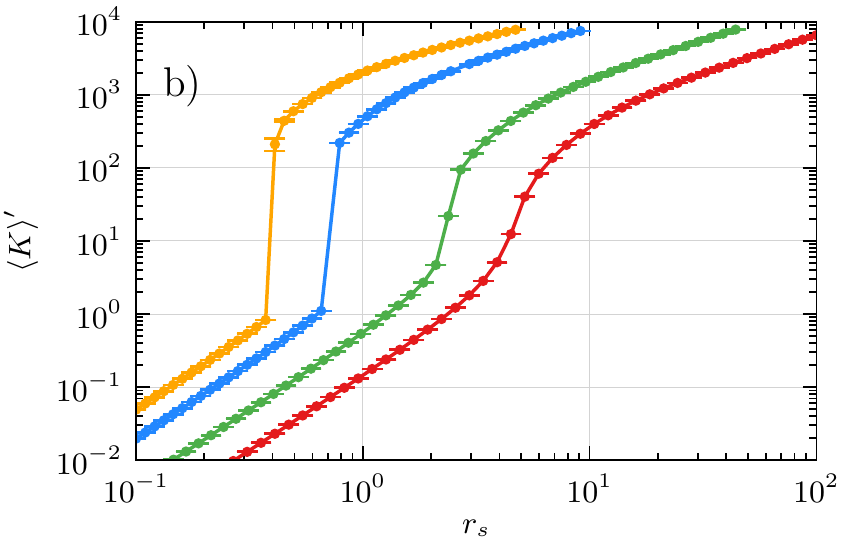}
\caption{(color online). Average sign \textbf{a)} and average number of kinks \textbf{b)} of direct CPIMC, plotted versus the density parameter for different particle numbers in $N_B=2109$ basis functions at $\theta=0.125$.}
\label{fig:avg_sign}
\end{figure}
Fig.~\ref{fig:avg_sign} a) shows the dependency of the average sign in CPIMC simulations of the UEG on the density parameter at a fixed degeneracy parameter $\theta=0.125$ for different particle numbers. The number of basis functions is fixed to $N_B=2109$, which is sufficient to obtain converged results (within reasonable statistical errors) for all data points. We generally observe a rather sharp drop of the average sign from almost $1$ to about $10^{-3}$. This effect clearly increases and shifts towards smaller $r_s$ with particle number. Consequently, for $N=33$ particles at this temperature we obtain negligible small statistical errors for $r_s\lesssim 0.4$, whereas for slightly larger values of $r_s$ direct simulations are not feasible. To investigate this behavior in more detail, in Fig.~\ref{fig:avg_sign}~b) we plot the average number of kinks in the simulations for the same parameters. This quantity is closely connected to the average sign since each additional kink in the paths comes with three potential sources of sign changes, cf.\ Sec.~\ref{sec:path}. In CPIMC simulations with on average more than $30$ kinks we find that, depending on the temperature, the average sign is too small to obtain results with reasonable statistical errors. 

In the high density regime, the average number of kinks grows linearly with $r_s$, see Fig.~\ref{fig:avg_sign}~b), then at some critical value of $r_s$ it starts growing exponentially. The slope of this exponential growth increases with particle number so that for $N=33$ it appears to be rather a jump from below $1$ to about $200$ kinks at $r_s\sim 0.4$ explaining the sudden drop of the average sign in Fig.~\ref{fig:avg_sign}~a). Interestingly, for further reduced density, the average number of kinks grows again linearly with $r_s$. We have carefully checked that this is not an effect of the finite number of basis functions. However, in this regime, even for $N=4$ particles the average number of kinks is larger than $1000$ resulting in a practically vanishing average sign. 
\begin{figure}[t]
\includegraphics{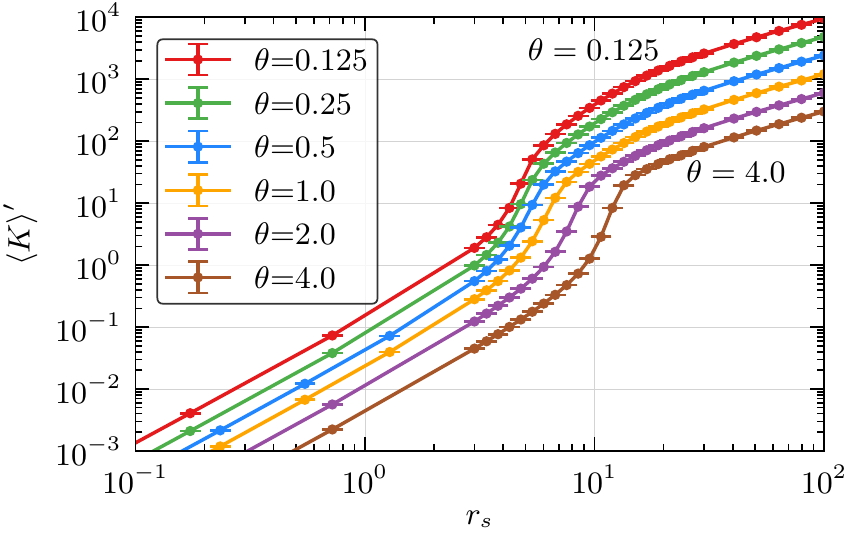}
\caption{(color online). Average number of kinks of direct CPIMC, plotted versus the density parameter for $N=4$ particles in $N_B=5575$ basis functions at different temperatures. }
\label{fig:avgKink_diff_temp}
\end{figure}
For $N=4$ particles, Fig.~\ref{fig:avgKink_diff_temp} shows the average number of kinks in dependence on $r_s$ for different degeneracy parameters. In the linear regimes (both at very large and small values of $r_s$), the average number of kinks depends also linearly on the degeneracy parameter while the onset of the exponential growth shifts towards smaller $r_s$, for increasing degeneracy, i.e.\ for decreasing $\theta$. Further, the transition from the exponential to the linear $r_s$ dependency is smoother the lower is the temperature, cf.\ red and brown curve in Fig.~\ref{fig:avgKink_diff_temp}. Summarizing, the direct CPIMC method suffers an abrupt drop of the average sign in particular for larger systems and lower temperature caused by a strong increase of the average number of kinks in the simulated paths.
\begin{figure}[t]
\includegraphics{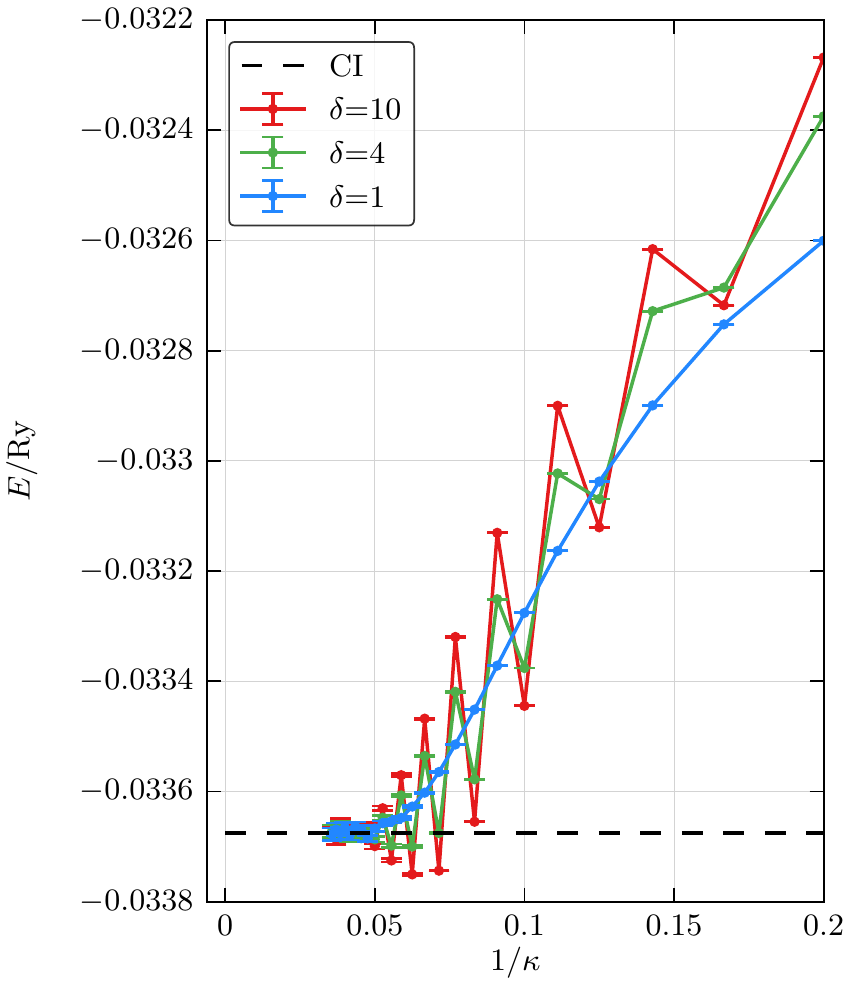}
\caption{(color online). Convergence of the internal energy with respect to the kink potential parameter $\kappa$, using different parameters $\delta$. The system consists of $N=4$ particles in $N_B=19$ basis functions at $\theta=0.5$ and $r_s=40$ for which the energy can be computed with an exact configuration interaction (CI) method (dashed black line). Each point is the result of a whole CPIMC simulation, where integer numbers from $5$ to $28$ have been used for $\kappa$.}
\label{fig:demonstrate_kinPot}
\end{figure}

\subsection{Extending CPIMC towards lower density\label{sec:kinkPot}}
In this section, the use of the auxiliary kink potentials is explained, and its influence on the CPIMC method is investigated in detail. These kink potentials have been introduced in~\cite{tim_prl15} to obtain the results for $r_s> 0.4$.

The average number of kinks in the simulation is only connected to the number of kinks $K$ necessary for the partition function of the primed system to be converged, cf.\ Eq.~(\ref{eq:primed_system}). However, to obtain correct physical observables via Eq.~(\ref{eq:average}) it is sufficient to include only those paths in the simulation that actually contribute to the physical partition function Eq.~(\ref{eq:Z_expansion}), which, due to cancellations of contributions with opposite sign, may converge for a much smaller value of $K$ than the primed partition function. In other words, if this cancellation applies, then we can restrict the simulation paths to a certain number of kinks and thereby strongly reduce the sign problem while still obtaining exact results for the observables. In addition, since both, Eq.~(\ref{eq:Z_expansion}) and~(\ref{eq:primed_system}), are  exact perturbation series in orders of the number of kinks $K$, it is reasonable to investigate the convergence of this series with respect to $K$. For this purpose, we have introduced an auxiliary Fermi-like kink potential
\begin{align}
V_{\delta,\kappa}(K)=\frac{1}{e^{-\delta(\kappa-K+0.5)}+1}\;,
\label{eq:fermiPot}
\end{align}
which becomes a step function at $K=\kappa+0.5$ in the limit $\delta \to \infty$. We add this potential as an auxiliary factor in the primed partition function so that it acts as a penalty, depending on the values of $\delta$ and $\kappa$, for paths with a large number of kinks. Hence, the simulated partition function is now parametrized by $\delta$ and $\kappa$ reading
\begin{align}
Z'(\delta,\kappa)=& \sum_{K=0,\atop K \neq 1}^{\infty} \sum_{\{n\}} \sum_{s_1\ldots s_{K-1}}\int\limits_{0}^{\beta} d\tau_1 \int\limits_{\tau_1}^{\beta} d\tau_2 \ldots \int\limits_{\tau_{K-1}}^\beta d\tau_K\,\label{eq:primed_system_potential} \\ \nonumber 
&V_{\delta,\kappa}(K)|W(K,\{n\}, s_1,\ldots,s_{K-1},\tau_1,\ldots,\tau_K)| \;.
\end{align} 
Obviously, for any non-negative value of $\delta$, we recover the original primed partition function in the limit $\kappa\to\infty$
\begin{align}
Z'=\lim_{\kappa \to \infty} Z'(\delta,\kappa)\;,\quad \forall\; \delta \geq 0\;. 
\end{align}
Therefore, performing CPIMC simulations for different values of $\kappa$ at fixed $\delta$ converges to the exact result in the limit $1/\kappa\to 0$. 

This is demonstrated in Fig.~\ref{fig:demonstrate_kinPot}, where the convergence of the internal energy is shown for three different values of $\delta$. The system size has been chosen to be very small, i.e.\ $N=4$ particles in $N_B=19$ basis functions at $\theta=0.5$ and $r_s=40$, so that the energy can be computed with an exact diagonalization method (dashed black line). For the parameter $\kappa$ integer values have been used from $\kappa=5$ to $28$. At $\delta=10$ (red points), the kink potential practically resembles a step function restricting paths in the simulation to a maximum of $K_{\text{\tiny max}}=\kappa$ kinks. Interestingly, in this case the energy converges not monotonically towards the exact result but oscillates with even and odd numbers of $\kappa$. Strictly speaking, for only odd numbers of $\kappa$ the energy does converge monotonically while for even numbers it first drops below the exact value before eventually converging. This behavior may be explained by the factor $(-1)^K$ in the weight function, c.f.~Eq.~(\ref{eq:weight}), dominating the other two sign changing sources of the phase factor and the two-electron integrals. Nevertheless, these oscillations render a reliable extrapolation to the exact limit $1/\kappa\to 0$ difficult and hence, simply restricting the number of kinks is not a good choice. For smaller values of $\delta$ (green points in Fig.~\ref{fig:demonstrate_kinPot}) where we, to a larger extent, allow paths with a larger number of kinks than $\kappa$, the oscillations are significantly reduced. At $\delta=1$ (blue points), the oscillations finally vanish completely and the energy converges monotonically towards the exact result.    
In fact, we always observe an s-shaped convergence behavior with $1/\kappa$ for Fermi potentials with $\delta \lesssim 1$. This allows for a very robust extrapolation scheme to the exact result in the limit $1/\kappa\to \infty$ after the onset of convergence that is clearly indicated by the change in curvature (at $\kappa\sim 17$ in Fig.~\ref{fig:demonstrate_kinPot}). 
\begin{figure}[t]
 \includegraphics{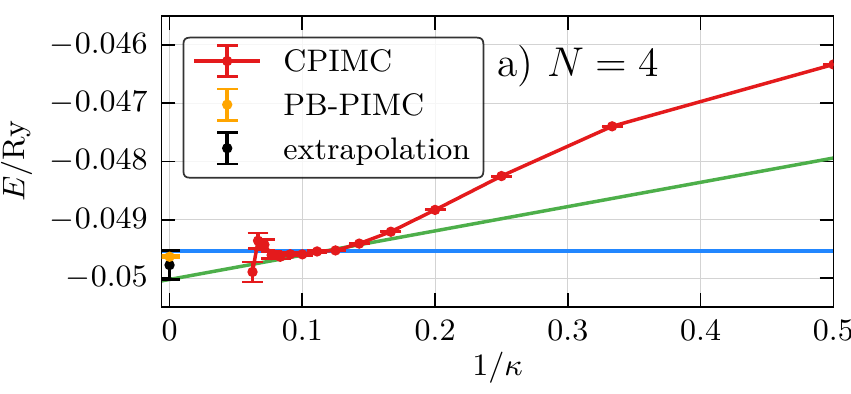}
  \includegraphics{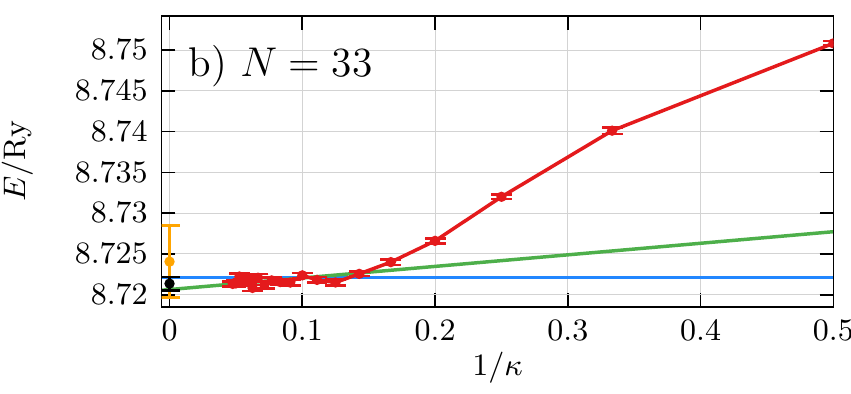}
 \caption{(color online). Convergence of the internal energy with respect to the kink potential parameter $\kappa$ and extrapolation to $1/\kappa \to 0$, corresponding to $K \to \infty$, at $\theta=1.0$.
\textbf{a)} $N=4$ particles and $r_s=10.0$ in $N_B=5575$ basis functions. \textbf{b)} $N=33$ and $r_s=1.0$ in $N_B=4169$ basis functions. The asymptotic values (black points) are enclosed between the blue and green lines and, within error bars, coincide with the PB-PIMC result (orange points).}
 \label{fig:kinkPot_fit}
\end{figure}

In Fig.~\ref{fig:kinkPot_fit}~a), we demonstrate this extrapolation scheme for a more difficult system of $N=4$ particles in $N_B=5575$ basis functions at $\theta=1$ and $r_s=10$, for which the direct CPIMC method without the kink potential is not applicable due to on average more than $50$ kinks, cf.~orange curve in Fig.~\ref{fig:avgKink_diff_temp}, and a resulting vanishing sign. To obtain an upper bound of the exact energy, we perform a horizontal fit (blue line) to those points after the onset of the convergence, while for the lower bound a linear fit is performed to those points (green line). The concrete fitting procedure is explained in appendix~\ref{fit_procedure}. For comparison the result for the energy of the likewise exact PB-PIMC method [cf. Sec.~\ref{sec:PB-PIMC}] is shown (orange point), which is well enclosed by the horizontal and linear fit and hence perfectly confirms our approach. Note that for the $N=4$ particles in only $N_B=19$ basis functions in Fig.~\ref{fig:demonstrate_kinPot} the energy is entirely converged for $\kappa=20$ so that all points for $\kappa>20$ lie on the horizontal line of the CI energy. This is because here the direct CPIMC algorithm converges to an average number of $20$ kinks. In contrast, in Fig.~\ref{fig:kinkPot_fit}~a), after the change in curvature at approximately $\kappa=8$, the energy is not entirely converged and still slowly decreasing. In this regime a near cancellation of all contributions for increasing $\kappa$ occurs. However, in the limit $\kappa\to \infty$ the energy does not converge linearly towards the exact value, because the direct CPIMC algorithm always converges at a finite value of $\braket{K}^\prime$, cf. Fig.~\ref{fig:avg_sign}~b) and Fig.~\ref{fig:avgKink_diff_temp}. Therefore, from some value of $\kappa$ onwards, depending on the average number of kinks in the direct CPIMC algorithm, the points will be on a horizontal line getting no further contributions for increasing $\kappa$. For this reason, the linear fit (green line) is indeed a true lower bound of the exact energy for the used number of basis function. Our extrapolation scheme also works well for larger systems, which is illustrated in Fig.~\ref{fig:kinkPot_fit}~b) for the example of $N=33$ particles at $\theta=1$ and $r_s=1$ in $N_B=4169$ basis functions. Here, the extrapolated value (black point) also agrees with the PB-PIMC result (orange point), which has a larger statistical error than in Fig.~\ref{fig:kinkPot_fit}~a), due to the larger density. For a convergence plot for the same system at a lower temperature of $\theta=0.0625$, where no other results are available, we refer to Ref.~\cite{tim_prl15}. 

In general, the use of the kink potential combined with the extrapolation scheme actually more  than doubles the accessible density parameter within the CPIMC approach at fixed other system parameters. Nevertheless, our procedure is still limited by the FSP, which is indicated by the increasing error bars of the last points in Fig.~\ref{fig:kinkPot_fit}~a). For example, at $\kappa=10$ there are on average $\braket{K}^\prime\sim 9.4$ kinks with a corresponding average sign $\braket{s}^\prime\sim 0.05$, while at $\kappa=16$ (last point) there are $\braket{K}^\prime\sim15.3$ kinks with a corresponding average sign $\braket{s}^\prime\sim 5\cdot10^{-3}$ causing a large statistical error. Of course, if the sign problem becomes too severe before the onset of convergence (indicated by the change in curvature), our procedure is not applicable.  

\section{Basic idea of PB-PIMC\label{sec:PB-PIMC}}
In contrast to CPIMC, our permutation blocking PIMC approach is essentially standard PIMC in coordinate space but combines two well-known concepts: 1) antisymmetric imaginary time propagators, i.e., determinants \cite{det1,det2,det3}, and 2) a fourth-order factorization of the density matrix \cite{chin1,chin2,sakkos}. In addition, to sample this more complicated configuration space, one of us has developed an efficient set of Monte Carlo updates based on the temporary construction of artificial trajectories. Since PB-PIMC and its application to the UEG have been introduced in detail in Refs.~\cite{pb_pimc_njp} and \cite{pb_pimc_jcp15}, here we shall restrict ourselves to a brief overview.

We start from the coordinate representation of the canonical partition function (\ref{eq:Z_trace}) describing a system of $N$ spin-polarized fermions at inverse temperature $\beta$
\begin{eqnarray}
 Z = \frac{1}{N!} \sum_{\sigma\in S_N} \textnormal{sgn}(\sigma) \int \textnormal{d}\mathbf{R}\ \bra{ \mathbf{R} } e^{-\beta\op{H}} \ket{ \op{\pi}_\sigma \mathbf{R}} \;, \label{boseZ}
\end{eqnarray}
with $\op{\pi}_\sigma$ being the exchange operator that corresponds to a particular element $\sigma$ from the permutation group $S_N$ with associated sign $\textnormal{sgn}(\sigma)$.
However, since the low-temperature matrix elements of $\op{\rho}$ are not known, we use the group property $\op\rho (\beta) = \prod_{\alpha=0}^{P-1} \op\rho(\epsilon)$, with $\epsilon=\beta/P$, and approximate each of the $P$ factors at a $P$ times higher temperature by the fourth-order factorization \cite{chin2,sakkos}
\begin{eqnarray}
 \label{cchin} e^{-\epsilon\op{H}} &\approx& e^{-v_1\epsilon\op{W}_{a_1}} e^{-t_1\epsilon\op{K}}\\ & & \nonumber e^{-v_2\epsilon\op{W}_{1-2a_1}} e^{-t_1\epsilon\op{K}} e^{-v_1\epsilon\op{W}_{a_1}} e^{-2t_0\epsilon\op{K}}\;,
\end{eqnarray}
which allows for sufficient accuracy, for small $P$.
The $\op{W}$ operators in Eq.\ (\ref{cchin}) denote a modified potential that combines the usual potential energy $\op{V}$ with
double commutator terms of the form 
\begin{eqnarray}
 [[\op{V},\op{K}],\op{V}] = \frac{\hbar^2}{m} \sum_{i=1}^N |\mathbf{F}_i|^2 \;, \quad \mathbf{F}_i = -\nabla_i V(\mathbf{R})\;,
\end{eqnarray}
where $\op{K}$ denotes the operator of the kinetic energy. Therefore, PB-PIMC requires the additional evaluation of all forces, and the final result for the partition function is given by
\begin{eqnarray}
\label{finalz} Z &=& \frac{1}{(N!)^{3P}} \int \textnormal{d}\mathbf{X} \prod_{\alpha=0}^{P-1} \Big( e^{-\epsilon\tilde V_\alpha}e^{-\epsilon^3u_0\frac{\hbar^2}{m}\tilde F_\alpha} \\ & & \textnormal{det}(\rho_\alpha)\textnormal{det}(\rho_{\alpha A})\textnormal{det}(\rho_{\alpha B}) \Big)\;.
\end{eqnarray}
Here, $\tilde V_\alpha$ and $\tilde F_\alpha$ contain all contributions of the potential energy and the forces, respectively, and the diffusion matrix is given by
\begin{eqnarray}
 \rho_\alpha(i,j) &=& \lambda_{t_1\epsilon}^{-D} \textnormal{exp} \left( -\frac{\pi}{\lambda^2_{t_1\epsilon}} ( \mathbf{r}_{\alpha,j} - \mathbf{r}_{\alpha A,i})^2 \right)\;, \label{diffusion}
\end{eqnarray}
with $\lambda_{t_1\epsilon}=\sqrt{2\pi\epsilon t_1\hbar^2/m}$ being the thermal
wavelength of a single ``time slice''.

Instead of explicitly sampling each permutation cycle, as in standard PIMC, we combine both positively and negatively signed configuration weights in the determinants, which leads to a cancellation of terms and, therefore, a significantly increased average sign in our simulations.
However, this ``permutation blocking'' is only effective when $\lambda_{t_1\epsilon}$ is comparable to the mean inter-particle distance.
With increasing $P$, $\lambda_{t_1\epsilon}$ decreases and the average sign eventually converges towards that of standard PIMC. 
Hence, it is crucial to combine the determinants with the fourth order factorization from Eq.\ (\ref{cchin}), which allows for sufficient accuracy with as few as two or three propagators and thereby maximizes the benefit of the blocking by the determinants.

\section{CPIMC and PB-PIMC benchmark results for the polarized UEG\label{sec:results}}
Due to the complementary character of the FSP the CPIMC and PB-PIMC approaches are well suited to be combined and, thereby, to circumvent the sign problem. Concerning the $N=33$ spin-polarized UEG, CPIMC is applicable practically over the entire temperature range from $\theta=0.01$ to $10$ and suffers an increasing sign problem for increasing $r_s$. The critical region at which the FSP becomes severe is around $r_s\sim 1$ for $\theta\lesssim 0.5$ and $r_s\sim \theta$ for $\theta\gtrsim 1$. On the other hand, the PB-PIMC method suffers a weak increase of the FSP for decreasing $r_s$, yet it is in principle capable of providing results over the entire density range for degeneracy parameters $\theta\gtrsim 0.75$. At temperatures $\theta < 0.5$, PB-PIMC is not feasible at high density. 
\begin{figure}[t]
\includegraphics{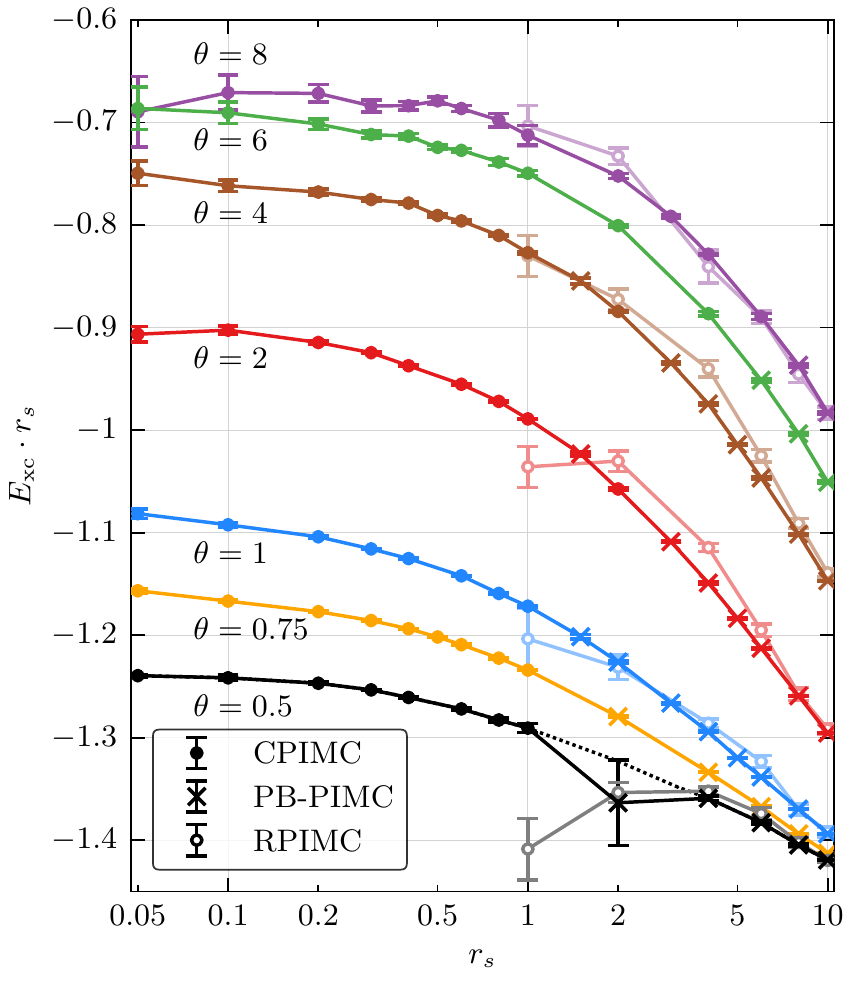}
\caption{(Color online) Exchange-correlation energy $E_{xc}$ times $r_s$ of the $N=33$ particle spin-polarized UEG over the density parameter $r_s$ for different degeneracy parameters $\theta$. Results have been obtained by combining the CPIMC (dots) and PB-PIMC (crosses) approach taking the most accurate values of each method (connected by the solid line). In addition, RPIMC results from~\cite{brown_prl13} are plotted for comparison (open circles).}
\label{fig:E_xc}
\end{figure}

\begin{figure}[t]
\includegraphics{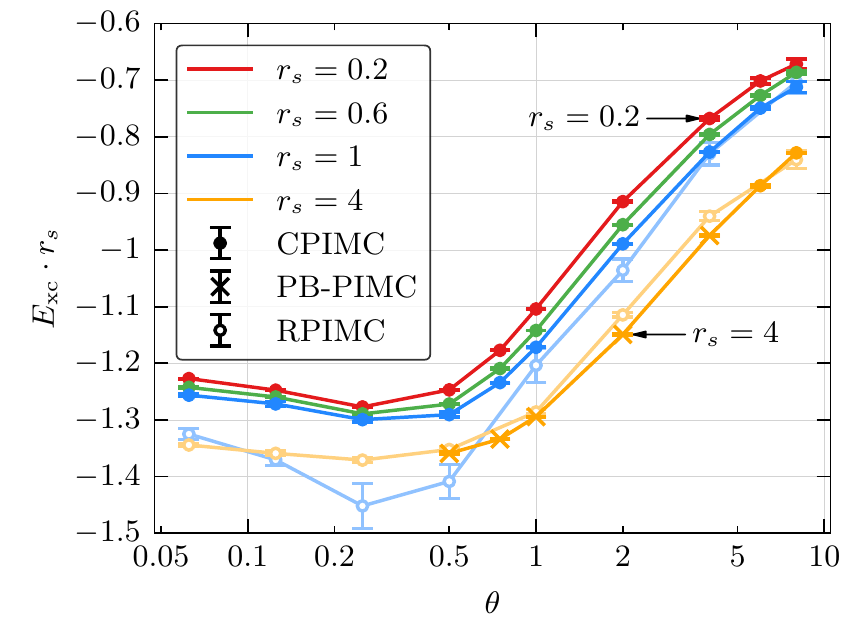}
\caption{(Color online) Exchange-correlation energy $E_{xc}$ times $r_s$ of the $N=33$ particle spin-polarized UEG over the degeneracy parameter $\theta$ for different density parameters $r_s$. Shown are results from CPIMC (dots) and PB-PIMC (crosses) calculations. In addition, RPIMC results from~\cite{brown_prl13} are plotted for comparison (lines with light colors and open circles, for $r_s=1$ and $r_s=4$).}
\label{fig:E_xc_theta}
\end{figure}

For the construction of density functionals the exchange-correlation energy $E_{xc}$ (per particle) of the UEG is of particular importance, which is obtained by subtracting the ideal energy $U_0$ from the total internal energy
\begin{align}
E_{xc}=E-U_0\;.
\end{align} 
In Fig.~\ref{fig:E_xc}, we show our results for the exchange-correlation energy. Note that we plot $E_{xc}\cdot r_s$ which converges towards the finite Hartree-Fock energy in the limit $r_s\to 0$. We always took the most accurate value of CPIMC (solid dots) or PB-PIMC (crosses), in cases where both are available. These data complement our earlier results that are included here as well, to have a complete set [for CPIMC, data for four isotherms $\theta=0.5,1,2,4$ have been reported in Ref.~\cite{tim_prl15}, while for PB-PIMC, the internal energy for the three isotherms $\theta=1,2,4$ has been presented in Ref.~\cite{pb_pimc_jcp15}, where the application of the method to the UEG is explained in detail]. At $\theta=0.5$, CPIMC can provide data up to $r_s=1$, while PB-PIMC suffers a too strong FSP below $r_s=2$ leaving a gap between both approaches. We have fitted a spline of order 4 to the available points and are thereby able to accurately close the gap (dotted line). With this, we are able to present {\em ab initio} results for this system for the entire density range, for all temperatures $\Theta \ge 0.5$.

In Tab.~\ref{tab:1}, we present all CPIMC and PB-PIMC data points shown in Fig.~\ref{fig:E_xc}. In addition to the exchange-correlation energy, the ideal, kinetic and potential energy are listed. Note that even the ideal energy in the canonical ensemble cannot be calculated analytically. Further, we added the number of basis functions $N_B$ that have been used in the corresponding CPIMC simulation, where we have carefully checked convergence of the energy (within statistical errors) with respect to $N_B$.
The origin of the fluctuations at the highest temperature are easily understood: 
 at $\theta=8$, the relative contribution of the exchange-correlation energy to the internal energy becomes very small since the kinetic energy dominates for increasing temperature. Hence, $E_{xc}$ is obtained by subtracting two large numbers of similar size which, of course, is ill-conditioned and, therefore, increases the statistical error of $E_{xc}$. The same applies in the limit $r_s\to 0$. Nevertheless, our  exchange-correlation energies represent the most accurate results published to date. 

For comparison we also plot the RPIMC data from Ref.~\cite{brown_prl13}. It is evident that they not only have a significantly larger statistical error, but they clearly deviate systematically from our results. Interestingly, the deviations increase from $\theta=1$ to $\theta=2$, and even at $\theta=4$, there is a significant discrepancy. This observation stands in contrast to the assumption that the systematic error due to the fixed node approximation vanishes for increasing temperature.  

Finally, Fig.~\ref{fig:E_xc_theta} shows the dependence of the exchange-correlation energy on temperature for four fixed densities. 
We again show the most accurate points of either CPIMC and PB-PIMC.
CPIMC allows for calculations practically down to the ground state, for $r_s\lesssim 1$. On the other hand, PB-PIMC is limited, at larger densities, to temperatures $\theta\geq 0.5$.
We observe that all isochores are nearly parallel and do not cross.
An interesting feature is the existence of a minimum around $\Theta\sim 0.25$, for all densities [some uncertainty remains for the lowest density, $r_s=4$, as our simulations are confined to $\Theta \ge 0.5$]. Similar observations have been made in the fit results of Ref.~\cite{karasiev_prl14} and in the computation of the screened potential of an ion in a streaming quantum plasma \cite{zhandos_pre_15}. 

The origin of this non-monotonic behavior is a competition of two effects. The governing trend is a decrease of the (modulus of the) interaction energy with temperature arising from a thermal broadening of the particle density. At low temperatures there exists a second trend arising from quantum diffraction effects: the thermal DeBroglie wavelength is reduced with temperature increase which increases the Coulomb interaction. A similar trend of an intermediate increase of correlations with temperature has been predicted for Wigner crystallization in 2D~\cite{ceperley_prl_09}.

In addition to the {\em ab initio data}, Fig.~\ref{fig:E_xc_theta} also includes the fixed node RPIMC data of Ref.~\cite{brown_prl13} which are available for the two lowest densities, $r_s=1$ and $r_s=4$. For the case $r_s=4$ the RPIMC results are systematically too high by a few percent. More severe deviations are observed for $r_s=1$ where the energies are too low. Particularly strong deviations are seen for low temperatures, $\theta \lesssim 1$ where the error exceeds $10\%$, giving even rise to a crossing of two isochores.

\section{Summary and Discussion}
This paper was devoted to a detailed discussion of the CPIMC simulation results for the uniform electron gas reported in a recent Letter \cite{tim_prl15}. We presented a systematic analysis of the fermion sign problem of direct CPIMC for the polarized UEG. For increasing particle number, a sharp drop of the average sign, at a certain critical value of $r^{\rm cr}_s(\Theta, N)$, has been observed and was shown to be connected to a strong increase in the average number of kinks in the simulation paths in Fock space. By introducing an auxiliary Fermi-like kink potential we introduced a modified CPIMC approach for which 
the accessible $r_s$-range could be increased by more than a factor $2$, for a fixed particle number and temperature \cite{tim_prl15}. When restricting the number of kinks to a maximum number $K_{\text{\small max}}$, it turned out that the energy does not converge monotonically but rather oscillates towards the exact result with increasing $K_{\text{\small max}}$, which renders a reliable extrapolation scheme difficult. However, by choosing the kink potential parameter $\delta$ such that it acts as a smooth penalty for paths with a larger number of kinks, a monotonic convergence of the energy could be achieved. We have developed a robust extrapolation scheme that provides  strict upper and lower bounds thereby yielding an accurate value for the thermodynamic quantities of the UEG. 

An independent confirmation of our extrapolation procedure could be obtained by a comparison to accurate PB-PIMC results. Interestingly, utilizing the kink potential, the energy of the simulation typically converges at about $20-30$ kinks (on average in the simulation paths), whereas the direct CPIMC approach (without the potential) equilibrates at several thousand kinks. This is explained by an almost complete cancellation of contributions of paths with a large number of kinks in the partition function, which sets the limitation of the auxiliary kink potential method: it works only if we are able to reach the onset of this near cancellation, before the sign problem becomes too severe. This is clearly detectable from the convergence behavior of the energy, cf. Fig.~\ref{fig:kinkPot_fit}: only when the energy approaches the horizontal asymptote, as a function of $1/\kappa$, the method is applicable.

The second goal of this paper was to extend the available {\em ab initio} results for the exchange-correlation energy of the polarized electron gas to higher temperatures and lower densities. This was achieved by combining two complementary independent methods---CPIMC and PB-PIMC. With this we were able to avoid the sign problem for $N=33$ electrons over the entire density range, for all temperatures  $\theta \ge 0.5$, and we presented data up to $\theta=8$, completely avoiding fixed nodes or similar approximations. In all cases where both methods overlap we observed perfect agreement (within error bars), allowing for extremely valuable cross-checks.

Below $\theta=0.5$, the present combination of two methods accesses only parts of the density range. Within the current implementations (and reasonable numerical effort) PB-PIMC is not applicable, for high densities, whereas CPIMC can only provide accurate results up to a minimum density around $r_s\sim 1$, leaving open a gap in the density which further increases with the particle number. Work is presently under way to access larger particle numbers and, eventually, perform an extrapolation to the thermodynamic limit, as was successfully demonstrated for very high densities in Ref.~\cite{tim_prl15}.

The present benchmark results should be useful for the development of improved quantum Monte Carlo simulations including density matrix QMC \cite{foulkes_14,foulkes_15} and tests of improved fermionic nodes for RPIMC. The present scheme of combining CPIMC and PB-PIMC should also be suitable to produce first-principle results for the paramagnetic electron gas for which an increased sign problem of CPIMC was observed~\cite{tim_prl15}. 

\section*{Acknowledgements}
This work is supported by the Deutsche Forschungsgemeinschaft via project BO 1366-10 and via SFB TR-24 project A9 as well as grant shp00015 for CPU time at the Norddeutscher Verbund f\"ur Hoch- und H\"ochstleistungsrechnen (HLRN).

\appendix
\section{Extrapolation  with respect to the number of kinks \label{fit_procedure}}
To obtain an upper bound for the energy from CPIMC calculations utilizing the kink potential (see e.g. Fig.~\ref{fig:kinkPot_fit}) a horizontal (constant) fit is performed as follows: First, all data points with a relative error exceeding $1\%$ are discarded defining a maximum value of $\kappa$, denoted $\kappa_{\text{\small max}}$ (minimum of $1/\kappa_{\text{\small max}}$), satisfying this condition. Second, all data points are up-shifted by $1\sigma$ standard deviation. Then, horizontal fits are performed to the next $6,7,8\ldots,n_h$ points with $\kappa < \kappa_{\text{\small max}}$, where we add additional points as long as these deviate no more than $4\sigma$ from the constant fit. This procedure ensures that we only fit to those points belonging to the onset of convergence (indicated by the change in curvature in Fig.~\ref{fig:kinkPot_fit}). 

A lower bound of the energy is obtained by starting with a linear fit to the last $n_h$ points with $\kappa < \kappa_{\text{\small max}}$. But instead of the prior up-shift of the data by $1\sigma$ we now perform a down-shift of the data points by $1\sigma$ prior to the fit. We proceed with adding points included in the linear fit as long as there are less than $3$ points deviating by $2\sigma$ and less than $1$ point deviating by $3\sigma$ from the fit. The lower bound of the energy is given by the lowest value of all linear fits at $1/\kappa = 0$. The result for the energy is then computed as the mean value of the lower and upper bounds with the error estimated (from above) as their difference.


\begin{longtable*}[c]{@{}rr
@{\hspace{15pt}}  
S[table-parse-only,  table-text-alignment = center, table-number-alignment = left]
@{\hspace{15pt}}                                                              
S[table-parse-only,  table-text-alignment = center, table-number-alignment = left]
@{\hspace{15pt}}                                                              
S[table-parse-only,  table-text-alignment = center, table-number-alignment = left]
@{\hspace{15pt}}                                                              
S[table-parse-only,  table-text-alignment = center, table-number-alignment = left]
@{\hspace{15pt}}                                                              
S[table-parse-only,  table-text-alignment = center, table-number-alignment = left]
@{}}
\caption{Energies per particle for $N=33$ polarized electrons: ideal energy, $U_0$, kinetic energy, $E_{\rm kin}$, potential energy, $E_{\rm pot}$ and exchange-correlation energy $E_\text{xc}$. An \enquote{a} marks CPIMC results that have been obtained by an extrapolation as explained in appendix~\ref{fit_procedure}. For these values, the error includes systematic effects. All other errors correspond to a $1\sigma$ standard deviation. A \enquote{b} marks results from PB-PIMC calculations. For CPIMC results, the number of basis functions $N_B$ is given in the last column.  Energies in units of Ryd.}
\label{tab:1}\\\\
\toprule
 {$\theta$} &     {$r_s$} &              {$U_0$} &                          {$E_\text{kin}$} &                          {$E_\text{pot}$} &                        {$E_\text{xc}$} &  {$N_B$}\\
\midrule
\endfirsthead
\caption[]{\textit{(continued).} Energies per particle for $N=33$ polarized electrons: ideal energy, $U_0$, kinetic energy, $E_{\rm kin}$, potential energy, $E_{\rm pot}$ and exchange-correlation energy $E_\text{xc}$. An \enquote{a} marks CPIMC results that have been obtained by an extrapolation as explained in appendix~\ref{fit_procedure}. For these values, the error given in parenthesis includes systematic effects. All other errors correspond to a $1\sigma$ standard deviation. A \enquote{b} marks results from PB-PIMC calcuations. For CPIMC results, the number of basis functions $N_B$ is given in the last column.  Energies in units of Ryd.}\\\\
\toprule
 {$\theta$} &     {$r_s$} &              {$U_0$} &                          {$E_\text{kin}$} &                          {$E_\text{pot}$} &                        {$E_\text{xc}$} &  {$N_B$} \\
\midrule\endhead
  0.50 &   0.05 &     2380.191(6) &          2376.036(25) &          -20.63427(16) &            -24.789(26) &   2109 \\
       &   0.10 &    595.0477(16) &           593.041(25) &          -10.40869(32) &            -12.416(25) &   4169 \\
       &   0.20 &     148.7619(4) &            147.818(5) &           -5.29077(12) &              -6.234(5) &   4169 \\
       &   0.30 &    66.11641(18) &           65.5186(17) &            -3.57994(9) &            -4.1777(17) &   4169 \\
       &   0.40 &    37.19048(10) &           36.7599(10) &           -2.72121(13) &            -3.1518(11) &   4169 \\
       &   0.60 &     16.52910(5) &   16.2673(14){${}^a$} &     -1.8577(8){${}^a$} &    -2.1198(21){${}^a$} &   2109 \\
       &   0.80 &    9.297620(25) &    9.1196(30){${}^a$} &      -1.424(4){${}^a$} &    -1.6034(26){${}^a$} &   2109 \\
       &   1.00 &    5.950477(16) &      5.823(6){${}^a$} &      -1.162(6){${}^a$} &      -1.291(4){${}^a$} &   2109 \\
       &   2.00 &     1.487619(4) &     1.426(22){${}^b$} &    -0.6202(23){${}^b$} &     -0.682(21){${}^b$} &        \\
       &        &                 &                       &                        &         -0.661{${}^c$} &       \\
       &   4.00 &   0.3719050(10) &     0.3618(6){${}^b$} &    -0.32970(8){${}^b$} &     -0.3398(5){${}^b$} &        \\
       &   6.00 &    0.1652910(5) &   0.16355(30){${}^b$} &    -0.22873(6){${}^b$} &   -0.23047(29){${}^b$} &        \\
       &   8.00 &  0.09297600(25) &   0.09356(14){${}^b$} &  -0.176150(30){${}^b$} &   -0.17557(13){${}^b$} &        \\
       &  10.00 &  0.05950500(16) &    0.06130(8){${}^b$} &  -0.143718(17){${}^b$} &    -0.14192(7){${}^b$} &        \\
  \midrule
  \pagebreak[3]
  0.75 &   0.05 &    3147.466(12) &            3143.18(4) &          -18.84333(19) &              -23.13(5) &   4169 \\
       &   0.10 &    786.8665(31) &           784.718(10) &            -9.51839(8) &            -11.667(11) &   4169 \\
       &   0.20 &     196.7166(8) &          195.6818(24) &            -4.85031(4) &            -5.8851(26) &   4169 \\
       &   0.30 &    87.42961(35) &           86.7672(12) &          -3.289850(30) &            -3.9523(12) &   4169 \\
       &   0.40 &    49.17916(19) &            48.7016(4) &          -2.506603(22) &             -2.9842(5) &   4169 \\
       &   0.50 &    31.47466(12) &          31.10585(31) &          -2.034685(20) &           -2.40349(34) &   4169 \\
       &   0.60 &     21.85740(9) &    21.5612(7){${}^a$} &   -1.71865(16){${}^a$} &    -2.0154(11){${}^a$} &   4169 \\
       &   0.80 &     12.29479(5) &    12.0878(5){${}^a$} &   -1.32039(10){${}^a$} &     -1.5280(8){${}^a$} &   4169 \\
       &   1.00 &    7.868665(31) &     7.7140(5){${}^a$} &     -1.0793(6){${}^a$} &     -1.2340(5){${}^a$} &   4169 \\
       &   2.00 &     1.967166(8) &     1.9097(6){${}^b$} &    -0.58218(7){${}^b$} &     -0.6397(6){${}^b$} &        \\
       &   4.00 &   0.4917920(19) &   0.47535(10){${}^b$} &  -0.316986(20){${}^b$} &   -0.33343(10){${}^b$} &        \\
       &   6.00 &    0.2185740(9) &   0.21257(13){${}^b$} &  -0.221880(28){${}^b$} &   -0.22788(13){${}^b$} &        \\
       &   8.00 &    0.1229480(5) &  0.120659(29){${}^b$} &  -0.171940(11){${}^b$} &  -0.174229(29){${}^b$} &        \\
       &  10.00 &  0.07868700(31) &  0.078268(32){${}^b$} &   -0.140854(9){${}^b$} &  -0.141272(31){${}^b$} &        \\
  \midrule
  \pagebreak[3]
  1.00 &   0.05 &    3957.262(19) &            3953.20(9) &          -17.56511(21) &              -21.63(9) &   4169 \\
       &   0.10 &      989.316(5) &           987.269(20) &           -8.87662(10) &            -10.923(21) &   4169 \\
       &   0.20 &    247.3289(12) &            246.337(5) &            -4.52798(5) &              -5.520(5) &   4169 \\
       &   0.30 &     109.9239(5) &          109.2790(18) &            -3.07450(4) &            -3.7194(19) &   4169 \\
       &   0.40 &    61.83222(30) &           61.3643(11) &          -2.345237(22) &            -2.8132(11) &   4169 \\
       &   0.60 &    27.48099(13) &            27.1891(4) &          -1.611535(18) &             -1.9034(4) &   4169 \\
       &   0.80 &     15.45805(8) &            15.2540(7) &            -1.2450(15) &             -1.4491(8) &   4169 \\
       &   1.00 &      9.89316(5) &    9.7381(10){${}^a$} &   -1.01625(29){${}^a$} &     -1.1717(7){${}^a$} &   4169 \\
       &   1.50 &    4.396958(21) &    4.3066(15){${}^b$} &   -0.71068(17){${}^b$} &    -0.8010(15){${}^b$} &        \\
       &   2.00 &    2.473289(12) &     2.4136(8){${}^b$} &   -0.55337(12){${}^b$} &     -0.6131(8){${}^b$} &        \\
       &   3.00 &     1.099239(5) &   1.06770(26){${}^b$} &    -0.39052(5){${}^b$} &   -0.42206(26){${}^b$} &        \\
       &   4.00 &   0.6183220(30) &   0.59980(14){${}^b$} &  -0.305012(33){${}^b$} &   -0.32353(15){${}^b$} &        \\
       &   5.00 &   0.3957260(19) &    0.38361(7){${}^b$} &  -0.251795(19){${}^b$} &    -0.26392(8){${}^b$} &        \\
       &   6.00 &   0.2748100(13) &    0.26690(5){${}^b$} &  -0.215138(13){${}^b$} &    -0.22305(5){${}^b$} &        \\
       &   8.00 &    0.1545810(8) &  0.150966(20){${}^b$} &   -0.167579(7){${}^b$} &  -0.171193(22){${}^b$} &        \\
       &  10.00 &    0.0989320(5) &  0.097380(13){${}^b$} &   -0.137806(5){${}^b$} &  -0.139358(13){${}^b$} &        \\
  \midrule
  \pagebreak[3]
  2.00 &   0.05 &      7335.15(4) &           7331.95(15) &          -14.93186(20) &             -18.13(15) &   5575 \\
       &   0.10 &    1833.788(11) &            1832.30(4) &           -7.54095(11) &               -9.02(4) &   5575 \\
       &   0.20 &    458.4470(26) &            457.718(8) &            -3.84305(5) &              -4.572(8) &   5575 \\
       &   0.30 &    203.7542(12) &          203.2810(32) &            -2.60821(4) &            -3.0815(34) &   5575 \\
       &   0.40 &     114.6118(7) &          114.2583(20) &          -1.989454(30) &            -2.3429(21) &   5575 \\
       &   0.60 &    50.93856(29) &            50.7147(9) &          -1.368155(19) &            -1.5920(10) &   5575 \\
       &   0.80 &    28.65294(16) &            28.4931(5) &          -1.055120(15) &             -1.2149(5) &   5575 \\
       &   1.00 &    18.33788(11) &          18.21454(30) &          -0.865709(17) &           -0.98905(31) &   5575 \\
       &   1.50 &      8.15017(5) &    8.0775(12){${}^b$} &   -0.60945(21){${}^b$} &    -0.6821(12){${}^b$} &        \\
       &   2.00 &    4.584470(26) &     4.5339(4){${}^a$} &     -0.4780(5){${}^a$} &     -0.5287(6){${}^a$} &   5575 \\
       &   3.00 &    2.037542(12) &   2.00917(27){${}^b$} &    -0.34120(8){${}^b$} &   -0.36958(28){${}^b$} &        \\
       &   4.00 &     1.146118(7) &   1.12840(27){${}^b$} &    -0.26956(9){${}^b$} &   -0.28727(28){${}^b$} &        \\
       &   5.00 &     0.733515(4) &    0.72143(9){${}^b$} &  -0.224617(35){${}^b$} &   -0.23670(10){${}^b$} &        \\
       &   6.00 &   0.5093860(29) &   0.50075(11){${}^b$} &    -0.19347(5){${}^b$} &   -0.20210(13){${}^b$} &        \\
       &   8.00 &   0.2865290(16) &    0.28185(4){${}^b$} &  -0.152706(18){${}^b$} &    -0.15739(4){${}^b$} &        \\
       &  10.00 &   0.1833790(11) &  0.180676(18){${}^b$} &  -0.126841(10){${}^b$} &  -0.129543(21){${}^b$} &        \\
  \midrule
  \pagebreak[3]
  4.00 &   0.05 &    14258.10(14) &          14256.29(19) &          -13.17459(10) &             -14.99(23) &  24405 \\
       &   0.10 &    3564.525(35) &            3563.55(5) &            -6.63750(5) &               -7.62(6) &  24405 \\
       &   0.20 &      891.131(9) &           890.660(12) &          -3.367889(23) &             -3.839(14) &  24405 \\
       &   0.30 &      396.058(4) &            395.752(5) &          -2.277115(17) &              -2.583(7) &  24405 \\
       &   0.40 &    222.7828(22) &          222.5676(30) &          -1.731134(13) &              -1.946(4) &  24405 \\
       &   0.50 &    142.5810(14) &          142.4029(24) &          -1.403167(13) &            -1.5812(27) &  24405 \\
       &   0.60 &     99.0146(10) &           98.8721(13) &           -1.184072(9) &            -1.3265(16) &  24405 \\
       &   0.80 &      55.6957(5) &           55.5925(13) &          -0.909464(11) &            -1.0126(14) &  24405 \\
       &   1.00 &    35.64525(35) &           35.5622(10) &          -0.743926(12) &            -0.8269(10) &  24405 \\
       &   1.50 &    15.84233(15) &   15.7935(18){${}^b$} &     -0.5208(4){${}^b$} &    -0.5696(19){${}^b$} &        \\
       &   2.00 &      8.91131(9) &           8.87718(18) &           -0.407967(8) &           -0.44210(20) &  24405 \\
       &   3.00 &      3.96058(4) &     3.9409(4){${}^b$} &   -0.29176(17){${}^b$} &     -0.3115(5){${}^b$} &        \\
       &   4.00 &    2.227828(22) &   2.21563(34){${}^b$} &   -0.23140(14){${}^b$} &     -0.2436(4){${}^b$} &        \\
       &   5.00 &    1.425810(14) &   1.41669(16){${}^b$} &    -0.19370(8){${}^b$} &   -0.20282(18){${}^b$} &        \\
       &   6.00 &    0.990146(10) &   0.98344(14){${}^b$} &    -0.16772(8){${}^b$} &   -0.17442(17){${}^b$} &        \\
       &   8.00 &     0.556957(5) &    0.55306(9){${}^b$} &    -0.13378(5){${}^b$} &   -0.13767(10){${}^b$} &        \\
       &  10.00 &   0.3564530(35) &    0.35389(4){${}^b$} &  -0.112127(25){${}^b$} &    -0.11469(4){${}^b$} &        \\
  \midrule
  \pagebreak[3]
  6.00 &   0.05 &    21232.56(31) &          21231.34(28) &           -12.50240(7) &               -13.7(4) &  38911 \\
       &   0.10 &      5308.14(8) &            5307.53(7) &            -6.28885(4) &              -6.90(11) &  38911 \\
       &   0.20 &    1327.035(19) &          1326.709(17) &          -3.181308(18) &             -3.507(26) &  38911 \\
       &   0.30 &      589.793(9) &            589.566(8) &          -2.145065(12) &             -2.372(12) &  38911 \\
       &   0.40 &      331.759(5) &            331.602(5) &           -1.626612(9) &              -1.783(7) &  38911 \\
       &   0.50 &    212.3256(31) &          212.1926(34) &           -1.315274(9) &              -1.448(5) &  38911 \\
       &   0.60 &    147.4484(21) &          147.3440(24) &           -1.107473(8) &            -1.2118(32) &  38911 \\
       &   0.80 &     82.9397(12) &             82.864(4) &          -0.847318(17) &              -0.923(4) &  38911 \\
       &   1.00 &      53.0814(8) &           53.0227(25) &          -0.690775(17) &            -0.7494(26) &  38911 \\
       &   2.00 &    13.27035(19) &            13.2448(6) &          -0.374712(10) &             -0.4003(6) &  38911 \\
       &   4.00 &      3.31759(5) &     3.3074(5){${}^a$} &    -0.21107(8){${}^a$} &     -0.2216(6){${}^a$} &  38911 \\
       &   6.00 &    1.474484(21) &   1.46893(21){${}^b$} &   -0.15299(14){${}^b$} &   -0.15854(26){${}^b$} &        \\
       &   8.00 &    0.829397(12) &   0.82618(12){${}^b$} &    -0.12223(8){${}^b$} &   -0.12544(15){${}^b$} &        \\
       &  10.00 &     0.530814(8) &    0.52859(9){${}^b$} &    -0.10284(7){${}^b$} &   -0.10507(11){${}^b$} &        \\
  \midrule
  \pagebreak[3]
  8.00 &   0.05 &      28224.1(5) &            28222.5(4) &           -12.14740(7) &               -13.8(7) &  44473 \\
       &   0.10 &     7056.03(14) &           7055.43(10) &          -6.103529(32) &              -6.71(17) &  44473 \\
       &   0.20 &    1764.009(34) &          1763.732(25) &          -3.081446(15) &               -3.36(4) &  44473 \\
       &   0.30 &     784.004(15) &           783.799(12) &          -2.073757(12) &             -2.279(19) &  44473 \\
       &   0.40 &      441.002(9) &            440.863(7) &           -1.569731(9) &             -1.709(11) &  44473 \\
       &   0.50 &      282.241(5) &            282.151(5) &           -1.267116(9) &              -1.358(8) &  44473 \\
       &   0.60 &      196.001(4) &          195.9224(35) &           -1.065252(7) &              -1.144(5) &  44473 \\
       &   0.80 &    110.2505(21) &            110.191(8) &          -0.812627(18) &              -0.872(8) &  44473 \\
       &   1.00 &     70.5603(14) &             70.509(9) &          -0.660769(30) &             -0.712(10) &  44473 \\
       &   2.00 &    17.64009(34) &           17.6191(11) &          -0.355004(10) &            -0.3760(12) &  44473 \\
       &   3.00 &     7.84004(15) &             7.8274(5) &           -0.251192(7) &             -0.2638(5) &  44473 \\
       &   4.00 &      4.41002(9) &           4.40107(15) &           -0.198143(7) &           -0.20709(17) &  44473 \\
       &   6.00 &      1.96001(4) &   1.95532(32){${}^a$} &    -0.14327(8){${}^a$} &     -0.1482(5){${}^a$} &  44473 \\
       &   8.00 &    1.102505(21) &   1.09990(18){${}^b$} &   -0.11447(13){${}^b$} &   -0.11708(22){${}^b$} &        \\
       &  10.00 &    0.705603(14) &   0.70369(11){${}^b$} &    -0.09639(7){${}^b$} &   -0.09830(13){${}^b$} &        \\
\bottomrule
\end{longtable*}

\end{document}